\documentclass{iaus}
\usepackage{graphicx}

\title[Simulations of surface convection in low-mass stars]
{Radiation-hydrodynamics simulations of surface convection in low-mass stars:
  connections to stellar structure and asteroseismology}

\author[Ludwig, Caffau, \&\  Ku\v{c}inskas]   
{Hans-G.~Ludwig$^{1,2}$, Elisabetta~Caffau$^1$ \and A. Ku\v{c}inskas$^3$}

\affiliation{$^1$Observatoire de Paris-Meudon, GEPI, 92195 Meudon Cedex, France,
\break email: [Hans.Ludwig,Elisabetta.Caffau]@obspm.fr\\[\affilskip]
$^2$CIFIST Marie Curie Excellence Team, Observatoire de Paris-Meudo\\[\affilskip]
$^3$Institute of Theoretical Physics and Astronomy, Go\v {s}tauto 12, Vilnius
01108, Lithuania\break email: ak@itpa.lt}

\pubyear{2008}
\volume{252}  
\pagerange{1--6}
\date{?? and in revised form ??}
\jname{The art of modelling stars in the 21$^\mathrm{st}$ century}
\editors{L. Deng, K.L. Chan \& C. Chiosi, eds.}

\newcommand{\pun}[1]{\mbox{\rm\,#1}} 

\newcommand{\logg}{\ensuremath{\log g}}

\newcommand{\mlp}{\ensuremath{\alpha_{\mathrm{MLT}}}}

\newcommand{\moh}{\ensuremath{[\mathrm{M/H}]}}
\newcommand{\senv}{\ensuremath{\mathrm{s}_{\mathrm{env}}}}

\newcommand{\smin}{\ensuremath{\mathrm{s}_{\mathrm{min}}}}

\newcommand{\Teff}{\ensuremath{T_{\mathrm{eff}}}}

\newcommand{\COBOLD}{{\sf\mbox{CO$^5$BOLD}}}
\newcommand{\ttau}{\ensuremath{T(\tau)}}
\newcommand{\Pturb}{\ensuremath{P_\mathrm{turb}}}
\newcommand{\fturb}{\ensuremath{f_\mathrm{turb}}}
\newcommand{\vconv}{\ensuremath{v_\mathrm{c}}}

\begin{document}

\maketitle

\begin{abstract}
  Radiation-hydrodynamical simulations of surface convection in low-mass stars
  can be exploited to derive estimates of i) the efficiency of the convective
  energy transport in the stellar surface layers; ii) the convection-related
  photometric micro-variability. We comment on the universality of the
  mixing-length parameter, and point out potential pitfalls in the process of
  its calibration which may be in part responsible for the contradictory
  findings about its variability across the Hertzsprung-Russell digramme. We
  further comment on the modelling of the photometric micro-variability in
  HD\,49933 -- one of the first main \textit{COROT} targets.
\end{abstract}

\firstsection 
\section{Introduction}

Radiation-hydrodynamical (RHD) simulations of surface convection in low-mass
stars have reached a high level of maturity. Such simulations provide
quantitative predictions about the spatial and temporal statistics of the
flows taking place in the stellar surface layers. We used the 3D RHD code
\COBOLD\ to study convective flows in late-type stars at different
metalicity. This contribution deals with two distinct applications of
\COBOLD\ models. The first one is related to the efficiency of the convective
energy transport in convective envelopes, the second one to the low-level
photometric variability related temporal evolution of the surface granulation
pattern. We shall rather discuss problems than solutions with a focus on
dwarfs.

\section{Convective energy transport in the envelopes of late type-dwarfs}

It is well known from the theory of stellar structure that convection is
generally an efficient means of transporting energy, and it establishes a
thermal structure close to adiabatic. Only in vicinity of the boundaries of
convective regions noticeable deviations from adiabaticity occur.  In
convective envelopes of late-type stars the upper boundary of the convective
envelope -- usually located close to or even in the optically thin layers --
constitutes the bottle-neck for the energy transport through the stellar
envelope assigning a special role to it. Despite it is geometrically thin and
contains little mass it largely determines the properties of the convective
envelope as a whole. It is the value of the entropy of the adiabatically
stratified bulk of the convective~\senv\ which is most important from the
point of view of stellar structure since it influences the resulting radius
and effective temperature of a stellar model. \senv\ is controlled by the
efficiency of convective and radiative energy transport in the thin,
superadiabatically stratified surface layers.  Detailed RHD simulations can be
applied to model this region allowing to quantify the mutual efficiency of the
convective and radiative energy transport and predict \senv.  Comparing the
RHD predictions to standard 1D models based on mixing-length theory (MLT) the
value of \senv\ can be translated into a corresponding mixing-length
parameter~\mlp\ (see \cite[Ludwig et al. 1999]{ludwig99} for details).

In stellar evolution calculations the free mixing-length parameter is usually
calibrated against the Sun. However, it is unclear whether mixing-length
theory provides a suitable scaling of the convective efficiency at constant
\mlp\ across the Hertzsprung-Russell diagram (HRD), and a lot of work was
invested to address this issue empirically. Unfortunately, hitherto, no
coherent picture emerged. Here we report on an update of earlier work on the
theoretical calibration of \mlp\ based on RHD simulations. While not at all
comprehensive it illustrates some of the pitfalls in the process which may in
part be responsible of the blurred and sometimes even contradictory picture
which emerged so far concerning the variability of the mixing-length
parameter.

\subsection{Comments on the functional dependence of \mlp\ on stellar
parameters}

Discussions which took place during the symposium led the authors to add a
comment about the question which stellar properties govern the value of the
mixing-length parameter~\mlp. We argued above that RHD models of the surface
layers are able to provide information about \senv\ and \mlp. The models are
characterised by the {\em atmospheric parameters\/} and consequently the
functional dependence of \mlp\ can be described in terms of them.  Whether the
standard atmospheric parameters \Teff\ and \logg\ together with the chemical
composition are the most suitable coordinates is not clear.  One might
speculate that, e.g., the surface opacity is a physically more relevant
quantity.  Nevertheless, for unevolved late-type stars the conditions at the
stellar surface govern the global envelope structure and the standard
atmospheric parameters are suitable coordinates to parameterise them.  Global
stellar parameters (mass, radius, or age) play merely an implicit role.  The
situation only changes when the size of the granular cells or the thickness of
the superadiabatic layer become comparable to the stellar radius which might
happen in giants. To our opinion one should usually avoid to express changes
of \mlp\ in terms of, e.g., stellar mass or age since it tends to obscure the
underlying physics.

\subsection{\mlp\ from six 3D RHD models}

\begin{figure}
\begin{center}
\includegraphics[width=0.65\textwidth]{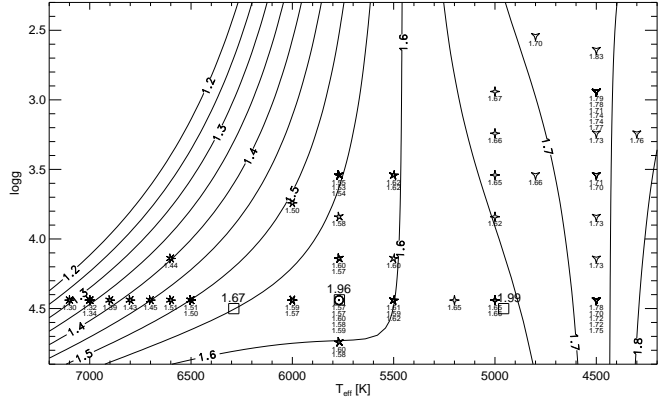}
\end{center}
\caption{Theoretically calibrated mixing-length parameters at solar
  metalicity in the \Teff-\logg-plane.  Squares mark our new results based on
  3D models, the other symbols mark earlier 2D results. The obtained values
  are given by the numbers, the isolines represent a smooth fit to the 2D
  data.}
\label{f:alpham00}
\end{figure}

\begin{figure}
\begin{center}
\includegraphics[width=0.65\textwidth]{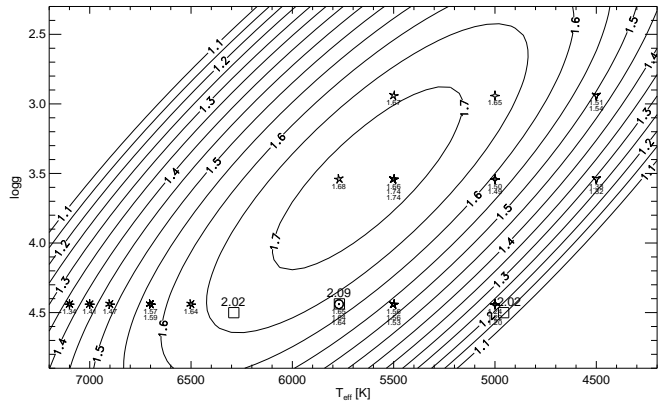}
\end{center}
 \caption{Same as Fig.~\ref{f:alpham00} but for $1/100$ of the solar metalicity.}
\label{f:alpham20}
\end{figure}

Squares in Figs.~\ref{f:alpham00} and~\ref{f:alpham20} mark the parameter
combinations in the \Teff-\logg-plane of six 3D RHD models we used for
calibrating \mlp. All atmospheres belong to dwarfs, three of solar
metalicity, three of a metalicity of $1/100$ solar. Our new 3D results are
superimposed on earlier results obtained from a grid of 2D RHD models by
\cite{ludwig99} and \cite{freytag99}. There are obvious differences of up to
$\approx 0.8$ between the 3D and 2D calibrated values. Does this indicate that
a theoretical calibration with the help of RHD models is hopelessly
inaccurate?

The first question to answer is what is a big and what is a small difference
when it comes to \mlp. At present interferometry and analysis of eclipsing
binaries provide stellar radii to an accuracy of about 1\,\%. Looking at
stellar structure models (see, e.g., \cite[Lebreton et al. 2001]{lebreton01}, their
Fig.~10) for late-type main-sequence stars the highest sensitivity of the
stellar radius and effective temperature to changes of \mlp\ is found at about
one solar mass. \cite{christensen97} finds for the Sun a sensitivity of
$\delta\ln R \approx -0.24\delta\ln\mlp$. Hence, considering the present
observationally achievable accuracies we would like to know \mlp\ to better
than 4\,\%, and the differences between the 2D and 3D results are clearly
relevant.

We did not mention yet the methodological changes we introduced when
calibrating our 3D-based \mlp's in comparison to the earlier works: i) Our new
values were calculated with a different mixing-length dialect. The earlier
results assumed the formulation given by \cite{boehmvitense58} while we now
used the formulation given by \cite{mihalas78}. ii) Following the general
trend in stellar evolution theory, our 1D comparison models are now
full-fledged stellar atmosphere models instead of integrating a prescribed
\ttau-relation to describe the atmospheric temperature run as was done in the
earlier results. As we shall see in a moment this is a crucial point, in
particular for the cool, metal-poor model at 5000\pun{K}. iii) It is now well
established that the convective energy transport operates more efficiently in
3D than in restricting 2D symmetry.  Hence, we would expect a
systematic bias towards higher mixing-length parameters in 3D relative to 2D.

Having in mind that we expect larger \mlp-values in 3D we think that the
absolute differences are perhaps not surprising and now rather focus on
differential trends with \Teff. The statistical uncertainties in the 2D
calibration amount to $\pm 0.05$. Taking together the different flavours of
MLT, and the different 1D comparison models used in the 2D and 3D calibrations
the trends start to look similar in 2D and 3D with the exception of the
coolest metal-poor model which is far off. What is the reason?

\begin{figure}
\begin{center}
\includegraphics[height=0.6\textwidth,angle=90]{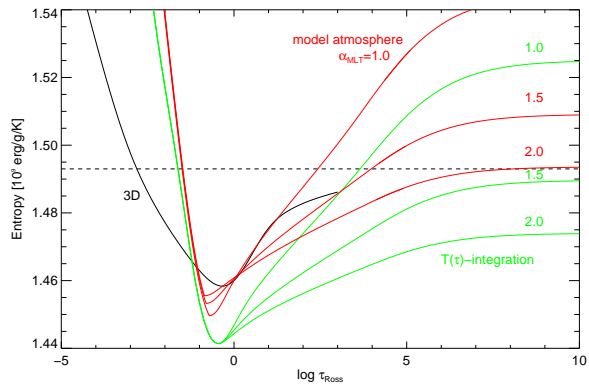}
\end{center}
\caption{Entropy profiles for 1D stellar atmosphere and stellar structure
  models for various mixing-length parameters in comparison to the mean 3D
  stratification for the metal-poor model at \mbox{\Teff=5000\pun{K}},
  \mbox{\logg=4.5}, and \mbox{\moh=-2.0}. All 1D models employ the MLT
  formulation by \cite{mihalas78}. The dashed line depicts the entropy~\senv\
  predicted by the 3D model for the adiabatically stratified part of the
  convective envelope.}
\label{f:entropymm20}
\end{figure}

Figure~\ref{f:entropymm20} shows a comparison between entropy profiles of 1D
MLT based models and of the horizontally and time-wised (on surfaces of equal
optical Rosseland depth) averaged 3D model. This kind of comparison provides
the calibrated \mlp. It is obvious that the result depends on the employed 1D
models. Using stellar atmosphere models produces a mixing-length parameter
which is almost by 0.5 larger than one based on models which use a prescribed
\ttau-relation in the convectively stable part of the stellar atmosphere. The
reason is to a large extend the difference in the entropy minimum~\smin\
attained in the deep atmospheric layer in combination with the small overall
entropy jump -- the difference $\senv-\smin$ -- in the star which amounts to
only $\approx 1/6$ of the solar value. Stars of higher \Teff\ show larger
entropy jumps which makes the exact level of the entropy minimum less
critical. The quite different differential behaviour of \mlp\ in 3D relative
to 2D is mainly a result of the different choice of comparison model.

While one might consider this as mere technicality we rather believe that part
of the confusion about trends and absolute values of the mixing-length
parameter have their origin -- besides observational problems -- in different
and not clearly specified procedures of how the MLT-related quantities are
computed and the atmospheric structure integration are actually performed in
models. It is, e.g., still quite common to use simple grey \ttau-relations for
describing the atmospheric temperature run. Figure~\ref{f:entropymm20}
illustrates that the particular choice might generate mismatches corresponding
to substantial changes in the value of the mixing-length parameter necessary
to restore the actual scaling of the envelope entropy with changing stellar
effective temperature.

Finally, Fig.~\ref{f:entropymm20} illustrates that even full-fledged model
atmospheres are not always able to reproduce the entropy minimum in 3D models
closely. The reason is that for metal-poor dwarfs 3D models predict large
deviations from radiative equilibrium conditions in the formally convectively
stable part of the atmosphere (in the figure apparent by the low entropy of
the 3D model at low optical depth). In 1D stellar atmospheresradiative
equilibrium conditions are assumed to fix the atmospheric temperatures.
\cite{trampedach07} puts forward the idea to extract the mean \ttau-relation
from 3D models for performing the atmospheric structure integration. This
obviously provides a better match to the 3D atmospheric entropy minimum, and
hence a ``cleaner'' calibration of the mixing-length parameter, however, also
is more difficult to implement in existing stellar evolution codes. The 3D
calibrated values of \mlp\ by Trampedach (for solar metalicity and MLT
formulation by B\"ohm-Vitense) are somewhat lower than our values presented
here but show a similar trend with \Teff\ on the main-sequence.

\subsection{Turbulent pressure trouble}

\begin{figure}
\begin{center}
\includegraphics[width=0.6\textwidth]{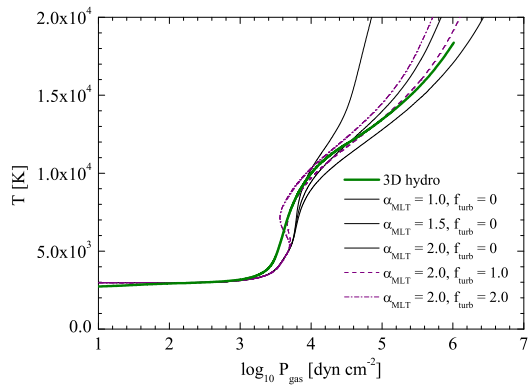}
\end{center}
\vspace{-\baselineskip}
  \caption{Temperature-pressure profiles of a 3D red giant model (thick solid
    line) in comparison to 1D stellar atmosphere models of different \mlp\
    leaving out (thin solid lines) or including (dashed and dashed-dotted
    line) turbulent pressure (for details see text).}
\label{f:pturb}
\end{figure}

In the main-sequence models discussed above turbulent pressure plays generally
only a small role but becomes relatively more important towards lower
gravities -- and causes extra trouble when one is interested in a well-defined
calibration of the mixing-length parameter. Figure~\ref{f:pturb} shows the
average temperature profile of a 3D red giant model (\mbox{\Teff=3600\pun{K}},
\mbox{\logg=1.0}, \mbox{\moh=0.0}) in comparison to standard 1D model
atmospheres of the same atmospheric parameters.  While turbulent
pressure~\Pturb\ is naturally included in the RHD simulation it is modelled in
a ad-hoc fashion in 1D models assuming a paramterisation
$\Pturb=\fturb\rho\vconv^2$, where \fturb\ is a free parameter of order unity,
$\rho$ the mass density and \vconv\ the convective velocity according MLT.

Figure~\ref{f:pturb} shows that in the 3D models the turbulent pressure
``lifts'' the temperature-gas pressure profile towards lower pressures which
is essentially impossible to reproduce in the 1D models -- irrespective of the
choice of \mlp\ and \fturb. The failure is related to the local nature of MLT
confining the action of the turbulent pressure gradients strictly to the
convectively unstable regions. While formally one can still match the thermal
profile of the 3D model in the deeper layers by 1D profile with suitably
chosen \mlp\ and/or \fturb\ such a match becomes physically little motivated
and is unlikely to provide a robust scaling with changing atmospheric
parameters. An improved 1D convection description including non-local effects
like overshooting is clearly desirable to handle this situation.  An empirical
calibration of \mlp\ using giants is likely to suffer from ambiguities related
to the way turbulent pressure is treated in 1D models.

\section{Modelling the photometric micro-variability in HD\,49933}

\begin{figure}
\begin{center}
\includegraphics[height=0.6\textwidth,angle=90]{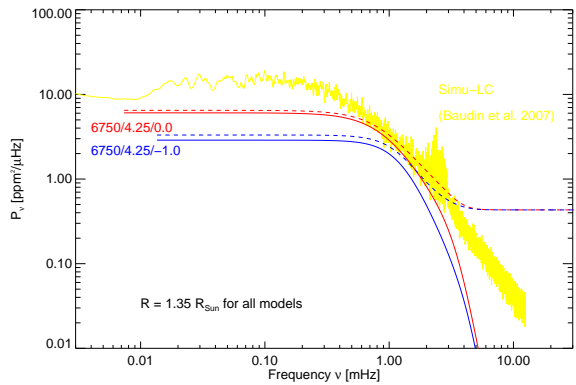}
\end{center}
\caption{Predictions for the temporal power spectral density of the
  photometric brightness fluctuations exhibited by HD\,49933. 3D RHD
  simulations are depicted by solid lines and labelled by
  \Teff/\logg/\moh. Dashed lines depict the result when adding the estimated
  photometric noise level. For further comparison a spectrum obtained with the
  \textit{COROT} light curve simulator (\cite[Baudin et al. 2007]{baudin07})
  is shown.}
\label{f:hd49933}
\end{figure}

The stochastically changing granular flow pattern on the surface of late-type
stars causes low-level residual brightness fluctuations in stellar
disk-integrated light. \cite{ludwig06} described a method to obtain
predictions of the observable temporal power spectrum from local-box RHD
simulations if -- besides the atmospheric parameters -- the radius of the
target star is known. One of the first primary asteroseismic targets of the
\textit{COROT} satellite mission was the metal-depleted
(\mbox{$-0.3\ge\moh\ge -0.4$}) F-dwarf HD\,49933 for which a
high-precision photometric light-curve was obtained. The data is not released
yet so that we are only showing theoretical predictions obtained from two RHD
simulation runs at \mbox{\Teff=6750\pun{K}}, \mbox{\logg=4.25} and two
metalicities \mbox{\moh=0.0 and -1.0} bracketing the observed metalicity of
HD\,49933 and having atmospheric parameters close to the spectroscopically
measured values. Figure~\ref{f:hd49933} depicts the power spectra obtained
from the two RHD simulations. For smoothing, the raw RHD spectra were fitted by
a simple analytical model, and oscillatory peaks from the acoustic eigenmodes
of the computational box were removed. Hence, the pure convection-related
signal is shown. The observed signal should fall between the two model
predictions, at least at around 1\pun{mHz} where the photon noise and the
signal due to magnetic activity are expected not to dominate. Unfortunately,
at present it looks that this is not quite the case. An actual comparison
will be presented in an upcoming paper by Ludwig \&\ Samadi, so stay tuned.

\end{document}